\documentclass[conference]{IEEEtran}

\IEEEoverridecommandlockouts
\usepackage{cite}
\usepackage{amsmath,amssymb,amsfonts}
\usepackage{graphicx}
\usepackage{textcomp}
\usepackage{xcolor}
\def\BibTeX{{\rm B\kern-.05em{\sc i\kern-.025em b}\kern-.08em
    T\kern-.1667em\lower.7ex\hbox{E}\kern-.125emX}}

\usepackage{booktabs} 
\usepackage{fancyvrb}
\usepackage{listings}
\usepackage{color}
\usepackage{balance}
\usepackage{fancybox}
\usepackage{xspace}
\usepackage{multirow}
\usepackage{url}
\usepackage{array}
\usepackage{graphicx}
\usepackage{bbding}
\usepackage{wasysym}
\usepackage{amssymb}
\usepackage{pdflscape}
\usepackage{tabularx,colortbl}
\usepackage{dirtree}
\usepackage{rotating}

\usepackage{comment}

\usepackage[noend]{algpseudocode}
\usepackage{wrapfig}

\usepackage{enumitem}

\usepackage{colortbl}
\usepackage{tabu}
\newcommand{\conclusion}[1]{\begin{center}\noindent\thicklines\setlength{\fboxsep}{4pt}\cornersize{0.03}\fbox{\begin{minipage}{3.36in}{\textbf{#1}}\end{minipage}}\end{center}}

\definecolor{Gray}{gray}{0.9}
\newcolumntype{g}{>{\columncolor{Gray}}l}

\definecolor{dkgreen}{rgb}{0,0.6,0}
\definecolor{gray}{rgb}{0.5,0.5,0.5}
\definecolor{mauve}{rgb}{0.58,0,0.82}

\begin{document}

\title{Studying the Role of Reusing Crowdsourcing Knowledge in Software Development}

\author{\IEEEauthorblockN{Rabe~Abdalkareem\\
}
	\IEEEauthorblockA{
		Department of Computer Science\\
		Faculty of Science\\
		Omar Al-Mukhtar University, Al Bayda, Libya\\
		Email: \url{rabe.abdalkareem@omu.edu.ly}}
}

\maketitle

\begin{abstract}
Crowdsourcing platforms, such Stack Overflow, has changed and impact the software development practice. In these platforms, developers share and reuse their software development and programming experience. Therefor, a plethora of research work focused on crowdsourcing in software engineering and showed that among other things, crowdsourcing development tend to increases developers’ productivity and reduce time-to-market. However, in crowdsourcing, the empirical studies of software quality are lacking and simple question such as what do developers use the crowdsourcing knowledge for is unanswered.

Therefor, our research focused on studying the impact of reusing crowdsourcing knowledge on software projects. To do so, we conduct several large scale empirical studies on some of the well-know crowdsourcing platforms including Stack Overflow and \textit{npm}. Our results showed that reusing knowledge from these crowdsourcing platforms have the potential to assist software development practice specifically in form of reusing crowd source code. However, using such a knowledge affects the quality of the software in several aspects such as makes the software projects suffer from dependency overhead and increases the maintenance effort. Based on these findings, we use the gained knowedlge to make sound data driven decision where we examine two software quality assurance methods to mitigate the risk of relying on crowd sourcing knowledge in software development. We examine the use of continuous integration (CI). Our analysis showed how CI can be improve to increase developers productivity and save their resources. 

\end{abstract}


\section{Research Problem and Hypothesis}
\label{sec:research_problem_and_hypothesis}
Crowdsourcing is emerging as an alternative outsourcing strategy which is gaining increasing attention in the software engineering. Amongst other things, crowdsourcing development tend to increases developers’ productivity and reduce time-to-market~\cite{mao2017survey, LaToza2016Software, Machado2014CrowdSoft}. Due to its advantages, the adapting of using crowdsourcing is growing rapidly. For example, an industrial report shows that the number of developers participating in software crowdsourcing increased 151\% from 2010 to 2011 and it is expected that the number of software companies developed using crowdsourcing will increase as well~\cite{massolution2012crowdsourcing}.

However, crowdsourcing software development involves complex tasks which differ significantly from the traditional outsourcing concept in software engineering. One main difference is that crowdsourcing are constructed as form of crowed knowledge on online sources such as Stack Overflow and GitHub~\cite{Stol2017Software}. That is, developers search these online sources and reuse this knowledge to address problems uncounted during software development. Additionally, reusing crowdsourcing knowledge is far from being a trivial task other type of reuse that known in software development such as using third part libraries— the latter are well established and studied practice, while crowd sourced knowledge is created and curated by a mostly uncoordinated collective contributions.

There has been a plethora of work focusing on crowdsourcing in software engineering and examine its benefits. For example, prior work examine crowd software testing \cite{Rojas2017ICSE,Mao2017ASE}, crowd software architecture design~\cite{LaToza2015ICSE}, and crowd requirement acquisition~\cite{Wang2014GSE}. However, in crowdsourcing, the empirical studies of software quality are lacking and simple question such as what do developers use the crowdsourcing knowledge for is unanswered.

Even though, the task of software crowdsourcing platforms is to provide a shared collaborative environment for software developers. These platforms structure the crowdsourcing in different forms. In this thesis, looked at two different crowdsourcing platforms based on their way of structuring the crowd knowledge. So, we destination between two types of software crowdsourcing knowledge; unstructured crowdsourcing knowledge and structured crowdsourcing knowledge. Although, both type of crowdsourcing knowledge depends on the crow (developers in our case) to construct the knowledge, they have their own challenges.

The main \textit{goal} of our research focus is to study the impact of using crowdsourcing knowledge in software engineering. First, following an empirical approach, we started by analyzing, and understanding and uncovering insights about crowdsourcing knowledge reused in software engineering. Then, we use the gained knowledge to make sound data driven decision and tool that software developers can use to develop and maintain their software projects. More specifically, in our PhD thesis, we made the following contributions:
\begin{itemize}[leftmargin=0.12in]
\item \textbf{\textit{Study the use of unstructured crowdsourcing knowledge on software quality.}} We first examine the use of unstructured crowdsourcing knowledge from Stack Overflow and found that software developers use crowdsourcing to gain knowledge about software development such as API usage and programming techniques~\cite{Abdalkareem2017software}. We also found that developers reuse source code snippets from Stack Overflow. We then empirically study Stack Overflow source code reused in mobile apps and found that code reuse from Stack Overflow tend to impact the software quality~\cite{Abdalkareem2017IST}.

\item \textbf{\textit{Examine structured crowdsourcing knowledge and its impact on software quality.}} The second part of this PhD investigates structure crowdsourcing platforms. We conducted an empirical study to examine the \textit{npm} platform~\cite{Abdalkareem2017FSE,Abdalkareem2017RC,Abdalkareem2018TSE}. We observed that JavaScript developers alleviate the lack of existing stander JavaScript library through the crowdsourcing constructed API and 16.8\% of the published packages on \textit{npm} tend to be trivial JavaScript packages. Additionally, we found that developer do no have an agreed opinion about these trivial packages. While some developers believe that trivial packages provide them with well-tested code, other tend to consider trivial packages increase the dependencies of their projects.

\item \textbf{\textit{Ensure the quality of the reused crowdsourcing knowledge.}} To ensure the quality of the reused crowdsourcing knowledge, we propose the use of software quality assurance methods. We start examining the use of contentious intonation (CI) development practice where we hypothesize that not every change to source code base of a project requires to be build~\cite{Abdalkareem2018CI,Abdalkareem2018CITSE}. 

\end{itemize}

To sum up, our thesis has an impact on the software engineering research community and practitioners through examining different crowdsourcing platforms and propose techniques to mitigate the impact of reusing such knowledge on software quality. Our results can help practitioners evolve guidelines and standards
for designing higher quality software and also understand deficiencies in current ways that developers reuse crowdsourcing knowledge.

The rest of the paper is organized as follows. Section~\ref{sec:thesis_contributions} presents the thesis contributions. Section~\ref{sec:thesis_contributions} summarizes the contributions of the thesis and identifies several avenues for future work in Section~\ref{sec:future_research_directions}. The conclusion is presented in Section!\ref{sec:conclusion}.

\section{ Thesis Contributions}
\label{sec:thesis_contributions}

The main contributions of my PhD thesis are divided into three parts. 
\subsection{Unstructured Crowdsourcing Knowledge.}
General crowdsourcing platforms such as Stack Overflow, heavily rely on contributions of the crowd to provide accumulated, quality knowledge to the software development community. Crowd knowledge on Stack Overflow is formated in unstructured way. Typically, users post questions on Stack Overflow, which are answered by one or (often) more participants. In essence, the job of answering the questions is outsourced to the crowd~\cite{LaToza2016Software}. However, the role of Stack Overflow has evolved to include much more than just answering questions. For example, Treude et al.~\cite{TreudeICSE2011} qualitatively analyzed a sample of Stack Overflow questions, and found that developers use Stack Overflow to share knowledge, provide development support, learn new technologies, and search for solutions to common and specific programming problems. However, aside from this qualitative study, almost no prior work exists that studied the reasons why developers use the crowdsourced knowledge on Stack Overflow for. Thus, we investigate and try to understand how developers actually use crowdsourced knowledge. More specifically, this research part aims to answer questions such as \textit{``what crowdsourced knowledge do developers actually use? And, what is the impact of reusing of such crowd knowledge on software quality?"}~\cite{Abdalkareem2017software}.

We took a different approach from previous studies and explored in what situations developers use the unstructured crowdsourcing knowledge available on Stack Overflow. Rather than mining Stack Overflow content, we mined GitHub commit histories looking for explicit mentions of Stack Overflow in their commit changes. We found that developers most often use the Stack Overflow to gain knowledge related to topics such as development tools, APIs usage, and operating systems. Additionally, we found that developers use Stack Overflow to promote Stack Overflow to other developers and providing rationale for a feature updates or additions. More important, we found that developers reuse source code snippets from Stack Overflow. 

Previous research showed that using crowdsourcing in software engineering development could negatively affect the quality of software system~\cite{LaToza2013CHI,StolICSE2014}. However, the impact of reusing source code from crowdsourcing platforms on software quality and maintenance remains to be addressed. Once the crowdsourcing knowledge has been integrated into a software system, it becomes a part of that software system. 

Thus, the goal of this part of our research is to study the impact of reusing crowdsourcing knowledge on the quality of software systems. More specifically, we perform an empirical study using 22 Android apps to investigate the impact of reusing crowdsourcing knowledge~\cite{Abdalkareem2017IST}. We found that the amount of reused Stack Overflow code varies for different mobile apps, that feature additions and enhancements are the main reasons for code reuse from Stack Overflow, mid-age and older apps reuse Stack Overflow code later in the lifetime and that more experienced developers reuse code in smaller teams/apps, while less experienced developers reuse code in larger teams/apps. More importantly, we examined the potential implications of code reuse from Stack Overflow and found that code reuse is related to higher potential of bug fixing commits. 
\conclusion{Our results shows that reusing unstructured crowdsourcing knowledge have the potential to assist software development practice. However, using such a knowledge may affect the quality of the software.}

\subsection{Structured Crowdsourcing Knowledge.}
Crowdsourcing is part of a wider practice of software development, shaped by the ability to collectively share developers' software development and programming experience in online code-sharing repositories such as GitHub~\cite{Stol2017Software}. Crowdsourcing in software engineering has led to the success of many forms, including open source systems’ development
(Linux and Apache)~\cite{LaToza2016Software} and code reuse (e.g,. \textit{npm}~\cite{npm11online}). An example of this open source software development is Node Package Management \textit{npm}. In \textit{npm}, the JavaScript developers community overcomes the lack of a rich stander JavaScript library through ``crowd API" where developers crowd the developing of JavaScript API. Unlike Stack Overflow platforms, the crowdsourcing knowledge generated on \textit{npm} is more structure. That is, the developers package JavaScript source code of functionalities and publish them as packages that other developers can use.

In this context of structured crowdsourcing knowledge, we conducted an empirical study to investigate the characteristic of such ``crowed generated API". More specifically, we examine one aspect of such crowd API that we call \textit{``trivial packages"}. We first define trivial packages as a package that contains code that a developer can easily code him/herself and hence, is not worth taking on an extra dependency for~\cite{Abdalkareem2017FSE, Abdalkareem2018TSE,Abdalkareem2017RC}. Like any other development practice, using trivial packages has its own proponents and opponents, so we set the goal to answer the question, \textit{``What are the reasons and drawbacks for developers to use trivial packages?"} We then employ qualitative and qualitative research methods to examine the prevalence, reasons and drawbacks of using trivial packages. We performed a survey with 88 JavaScript developers who use trivial packages and then we mine more than 230,000 \textit{npm} packages and 38,000 JavaScript projects. Our findings indicate that trivial packages are commonly and widely used in JavaScript projects. We also find that the majority of developers do not oppose the use of trivial packages and the main reasons developers use trivial packages is due to the fact that they are considered to be well implemented and tested. However, they do cite the fact that the additional dependencies’ overhead as a drawback of using these trivial packages. That said, our empirical study showed considering trivial packages to be well tested is a misconception since more than half of the trivial package we studied do not even have tests written, however, these trivial packages seem to be ‘deployment tested’ and have similar Tests, Community interest and Download count values as non-trivial packages. In addition, we find that some of the trivial packages have their own dependencies and, in our studied dataset, 11.5\% of the trivial packages have more than 20 dependencies. Hence, developers should be careful about which trivial packages they use.

\conclusion{Our findings illustrate that the use of crowdsourcing concept proves its self in mitigating the lack of stander JavaScript library. However, some challenges still exit for example the use of trivial packages.}

\subsection{Ensure the Quality of the Reused Crowdsourcing Knowledge.} 
The gained insights of the first two parts of our work provides us with direction to understand the impact of reusing crowdsourcing knowledge. And that existing research shows that software quality is one of the main concerns of adopting crowdsourcing development in software engineering. We believe that to improve the benefit and decrease the negative of reusing crowdsourcing knowledge on software systems, extra quality ensure steps are needed to be applied. While in traditional software development, a number of techniques have been used to ensure the quality of a software system, including testing~\cite{TSE2005Erdogmus}, code reviews~\cite{RigbyICSE2008}, and continuous integrations~\cite{VasilescuFSE2015}, in this part of our research, we hypothesize that traditional quality assurance techniques can be used to improve the adaption of crowdsourcing knowledge in software engineering through asking the question \textit{``Can traditional software quality assurance techniques mitigate the negative impact of reusing crowdsourcing knowledge?"} We aim at understanding to what extent existing quality assurance techniques can be applied as a means to mitigate the negative impact of reusing crowdsourcing knowledge on the quality of a software system and thus to improve and encourage the reuse of crowdsourcing knowledge.

To examine this hypothesis, we started by examine two quality assurance techniques. Since every reused code will be integrated to the software project through a commit, so we hypothesize that integrating the reused code is the right time to apply the quality assurance techniques. The idea is that when developers reuse crowdsourcing knowledge in form of source code, the the right time to control the quality of the reused code is in the time it is committed to the software projects. To start exploring this research area, we first examine the continuous integration~\cite{Abdalkareem2018CITSE}, \cite{Abdalkareem2018CI}.

Our main argument is that not every commit to the repository needs to trigger the CI process. For instance, developers may modify a project’s documentations and cause the CI process to be triggered. Since such a change does not affect the source code, the result of the build will not change and kicking off the CI process is just a waste of resources. Furthermore, in a discussion channel that is opened on Travis CI project, many developers argue that the CI process should not be run on every commit, and Travis CI developers are asked to provide an advanced mechanism to automatically CI skip
specific commits~\cite{Excludef56online_2}. Even though, Travis CI actually has built in functionality that allows developers to skip the CI process for a specific commit, the challenge of which commit to CI skip remains. As we will show later, developers often do not leverage this existing CI skip feature, which indicates that they a) either are unaware of this feature or b) do not know when a commit can be CI skipped. Therefore, the main goal is to identify the commit that is CI skip commit.

To address this research problem we developed two techniques. First we use rule-based technique based on manual analysis of more that 1,800 explicitly skip commit. The devised rule-based technique is able to detect and label CI skip commits with an AUC between 0.56 and 0.98 (average of 0.73). After that, we examine the use the use machine learning where we built a decision tree classifier based on 23 features extracted from historical data of software repositories. Our decision tree classifier achieves an average F1-measure of 0.79 (AUC of 0.89).

\conclusion{Our results shows that continuous integration auto-generation test case can be use to ensure the quality of the reused code from crowd sourcing platforms.}

\section{ Future Research Directions}
\label{sec:future_research_directions}
The long-term goal of our research is to leverage the developer's knowledge through crowdsourcing to help software developers produce high quality software and decrease maintenance effort of software projects. In this section, we high light some the future directions to achieve the aforementioned goal.

\subsection{Other Type of Crowdsourcing Knowledge.}
While this thesis has focused on the impact of reusing crowdsourcing knowledge on software quality in form of reusing source code, however, source code is not the only form of knowledge created by the crowd. For example, on Stack Overflow, developers generate crowdsourcing knowledge in form of discussion about development techniques. This knowledge is in free-text form or/and contain only code snippets that can be reuse to support software development decision for example design decisions. Recent work examine the use of textual crowdsourcing knowledge to recommend specific library for developers~\cite{Uddin2017ASE}. Another example of reusing non source code crowdsourcing knowledge is the use of apps' user reviews on app stores. In fact, there are several proposed approaches to help app developers improving their apps through using users' reviews to (e.,g.~\cite{palomba2018jss,TSEScalabrino2017,Palomba2015ICSME}). In future work, we plan to investigate the reuse of such non-source code knowledge in order to identify its impact. 

\subsection{More Realistic Evaluations.}
We have seen that reusing crowdsourcing knowledge affects the quality of software. However, our results is is based on analyzing open source projects. We believe that practitioners need to understand how this reusing crowdsourcing knowledge impact their projects qualities, what challenges arise when relying on such crowdsourcing knowledge. We plane to investigate study the impact reuse of crowdsourcing knowledge in an industrial setting.

\subsection{Understand the Encourage Participants.}
While this thesis focuses on the impact of reusing crowdsourcing knowledge on software quality, it also showed that one of the main factors that contributes to such impact the participants who generate the crowdsourcing knowledge. In fact, in a crowdsourcing platform, developers are voluntarily make some contributes. However, their willing to contribution is effect by several factors such as the type of the task that the participants can contribute to, the size of the task their background experiences. We plan to conduct a more in-depth studies to understand the real motivation for developers to contribute as well as the obstacle facing developers.

\subsection{Testing Generation for Source Code Snippets.}
In our work so far, we investigate the use of quality assurance to mitigate the risk of the reusing crowdsourcing knowledge. We found that CI process and auto-generation test cases can be use as a quality assurance technique. However, in our work, we do not consider the characteristic of source code generated by the crowd such as the context of the code. For example, on Stack Overflow, most of the posted source code is a snippet and it is not complete. We plan to further improve the auto-test generation techniques to be used on small incomplete source code snippets. 

\section{Conclusion}
\label{sec:conclusion}
The prevalence of crowdsourcing platforms (e.g., Stack Overflow, npm) has transformed software engineering by enabling developers to share and reuse code and experience, leading to increases in productivity and decreases in time-to-market. Despite these benefits, empirical understanding of the effects of this reuse on software quality remains incomplete. My work investigated the reuse of crowdsourced knowledge and examined its potential to assist software developers, particularly through code reuse. Crucially, I found that this practice tends to impact software quality by introducing dependency overhead and substantially raising maintenance effort. To address these risks, we utilized our findings to evaluate and propose two quality assurance mechanisms: optimizing continuous integration to save resources, and  to neutralize the negative consequences of reusing external knowledge.


\bibliographystyle{abbrv}
\balance
\bibliography{bibliography}

\begin{thebibliography}{10}

\bibitem{Abdalkareem2017RC}
R.~Abdalkareem.
\newblock Reasons and drawbacks of using trivial npm packages: The developers'
  perspective.
\newblock In {\em Proceedings of the 2017 11th Joint Meeting on Foundations of
  Software Engineering -Student Research Competition}, ESEC/FSE 2017, pages
  1062--1064. ACM, 2017.

\bibitem{Abdalkareem2018CI}
R.~Abdalkareem.
\newblock Predicting ci skip commits using machine learning.
\newblock In {\em The Conference adopting double-blined review process (To be
  Submitted)}, 2018.

\bibitem{Abdalkareem2018CITSE}
R.~Abdalkareem, S.~Mujahid, E.~Shihab, and J.~Rilling.
\newblock Which commits can be ci skipped? (submitted).
\newblock {\em IEEE Transactions on Software Engineering}, 2018.

\bibitem{Abdalkareem2017FSE}
R.~Abdalkareem, O.~Nourry, S.~Wehaibi, S.~Mujahid, and E.~Shihab.
\newblock Why do developers use trivial packages? an empirical case study on
  npm.
\newblock In {\em Proceedings of the 2017 11th Joint Meeting on Foundations of
  Software Engineering}, ESEC/FSE 2017, pages 385--395. ACM, 2017.

\bibitem{Abdalkareem2018TSE}
R.~Abdalkareem, V.~Oda, S.~Mujahid, and E.~Shihab.
\newblock Why do developers use trivial packages? an empirical case study on
  npm and pypi (to be submitted).
\newblock {\em IEEE Transactions on Software Engineering}, 2018.

\bibitem{Abdalkareem2017IST}
R.~Abdalkareem, E.~Shihab, and J.~Rilling.
\newblock On code reuse from stackoverflow: An exploratory study on android
  apps.
\newblock {\em Information and Software Technology}, 88:148 -- 158, 2017.

\bibitem{Abdalkareem2017software}
R.~Abdalkareem, E.~Shihab, and J.~Rilling.
\newblock What do developers use the crowd for? a study using stack overflow.
\newblock {\em IEEE Software}, 34(2):53--60, 2017.

\bibitem{TSE2005Erdogmus}
H.~Erdogmus, M.~Morisio, and M.~Torchiano.
\newblock On the effectiveness of the test-first approach to programming.
\newblock {\em IEEE Transactions on Software Engineering}, 31(3):226--237,
  2005.

\bibitem{FraserISSTA2013}
G.~Fraser, M.~Staats, P.~McMinn, A.~Arcuri, and F.~Padberg.
\newblock Does automated white-box test generation really help software
  testers?
\newblock In {\em Proceedings of the 2013 International Symposium on Software
  Testing and Analysis}, ISSTA 2013, pages 291--301. ACM, 2013.

\bibitem{Excludef56online_2}
P.~Ladenburger.
\newblock Exclude files from triggering a build · issue \#6301 ·
  travis-ci/travis-ci.
\newblock \url{https://github.com/travis-ci/travis-ci/issues/6301}, Jul 2016.
\newblock (accessed on 11/29/2017).

\bibitem{LaToza2015ICSE}
T.~D. LaToza, M.~Chen, L.~Jiang, M.~Zhao, and A.~v.~d. Hoek.
\newblock Borrowing from the crowd: A study of recombination in software design
  competitions.
\newblock In {\em 2015 IEEE/ACM 37th IEEE International Conference on Software
  Engineering}, volume~1, pages 551--562, 2015.

\bibitem{LaToza2013CHI}
T.~D. LaToza, W.~B. Towne, A.~van~der Hoek, and J.~D. Herbsleb.
\newblock Crowd development.
\newblock In {\em 2013 6th International Workshop on Cooperative and Human
  Aspects of Software Engineering (CHASE)}, pages 85--88, 2013.

\bibitem{LaToza2016Software}
T.~D. LaToza and A.~van~der Hoek.
\newblock Crowdsourcing in software engineering: Models, motivations, and
  challenges.
\newblock {\em IEEE Software}, 33(1):74--80, 2016.

\bibitem{Machado2014CrowdSoft}
L.~Machado, G.~Pereira, R.~Prikladnicki, E.~Carmel, and C.~R.~B. de~Souza.
\newblock Crowdsourcing in the brazilian it industry: What we know and what we
  don't know.
\newblock In {\em Proceedings of the 1st International Workshop on Crowd-based
  Software Development Methods and Technologies}, CrowdSoft 2014, pages 7--12.
  ACM, 2014.

\bibitem{mao2017survey}
K.~Mao, L.~Capra, M.~Harman, and Y.~Jia.
\newblock A survey of the use of crowdsourcing in software engineering.
\newblock {\em The Journal of Systems and Software}, 126(57-84), 2017.

\bibitem{Mao2017ASE}
K.~Mao, M.~Harman, and Y.~Jia.
\newblock Crowd intelligence enhances automated mobile testing.
\newblock In {\em 2017 32nd IEEE/ACM International Conference on Automated
  Software Engineering (ASE)}, pages 16--26, 2017.

\bibitem{massolution2012crowdsourcing}
Massolution.
\newblock {\em Crowdsourcing Industry Report: Enterprise Crowdsourcing: Market,
  Provider and Worker Trends}.
\newblock Massolution, 2012.

\bibitem{npm11online}
npm.
\newblock npm - documentation.
\newblock \url{https://www.npmjs.com/}.
\newblock (accessed on 06/27/2018).

\bibitem{palomba2018jss}
F.~Palomba, M.~Linares-V{\'a}squez, G.~Bavota, R.~Oliveto, M.~Di~Penta,
  D.~Poshyvanyk, and A.~De~Lucia.
\newblock Crowdsourcing user reviews to support the evolution of mobile apps.
\newblock {\em Journal of Systems and Software}, 137:143--162, 2018.

\bibitem{Palomba2015ICSME}
F.~Palomba, M.~Linares-Vásquez, G.~Bavota, R.~Oliveto, M.~D. Penta,
  D.~Poshyvanyk, and A.~D. Lucia.
\newblock User reviews matter! tracking crowdsourced reviews to support
  evolution of successful apps.
\newblock In {\em 2015 IEEE International Conference on Software Maintenance
  and Evolution (ICSME)}, pages 291--300, 2015.

\bibitem{RigbyICSE2008}
P.~C. Rigby, D.~M. German, and M.-A. Storey.
\newblock Open source software peer review practices: A case study of the
  apache server.
\newblock In {\em Proceedings of the 30th International Conference on Software
  Engineering}, ICSE '08, pages 541--550. ACM, 2008.

\bibitem{Rojas2017ICSE}
J.~M. Rojas, T.~D. White, B.~S. Clegg, and G.~Fraser.
\newblock Code defenders: Crowdsourcing effective tests and subtle mutants with
  a mutation testing game.
\newblock In {\em Proceedings of the 39th International Conference on Software
  Engineering}, ICSE '17, pages 677--688. IEEE Press, 2017.

\bibitem{TSEScalabrino2017}
S.~Scalabrino, G.~Bavota, B.~Russo, R.~Oliveto, and M.~D. Penta.
\newblock Listening to the crowd for the release planning of mobile apps.
\newblock {\em IEEE Transactions on Software Engineering}, pages 1--1, 2017.

\bibitem{Shamshiri2015ASE}
S.~Shamshiri, R.~Just, J.~M. Rojas, G.~Fraser, P.~McMinn, and A.~Arcuri.
\newblock Do automatically generated unit tests find real faults? an empirical
  study of effectiveness and challenges (t).
\newblock In {\em 2015 30th IEEE/ACM International Conference on Automated
  Software Engineering (ASE)}, pages 201--211, 2015.

\bibitem{StolICSE2014}
K.-J. Stol and B.~Fitzgerald.
\newblock Two's company, three's a crowd: A case study of crowdsourcing
  software development.
\newblock In {\em Proceedings of the 36th International Conference on Software
  Engineering}, ICSE 2014, pages 187--198. ACM, 2014.

\bibitem{Stol2017Software}
K.~J. Stol, T.~D. LaToza, and C.~Bird.
\newblock Crowdsourcing for software engineering.
\newblock {\em IEEE Software}, 34(2):30--36, 2017.

\bibitem{TreudeICSE2011}
C.~Treude, O.~Barzilay, and M.-A. Storey.
\newblock How do programmers ask and answer questions on the web? (nier track).
\newblock In {\em Proceedings of the 33rd International Conference on Software
  Engineering}, ICSE '11, pages 804--807. ACM, 2011.

\bibitem{Uddin2017ASE}
G.~Uddin and F.~Khomh.
\newblock Automatic summarization of api reviews.
\newblock In {\em Proceedings of the 32Nd IEEE/ACM International Conference on
  Automated Software Engineering}, ASE 2017, pages 159--170. IEEE Press, 2017.

\bibitem{VasilescuFSE2015}
B.~Vasilescu, Y.~Yu, H.~Wang, P.~Devanbu, and V.~Filkov.
\newblock Quality and productivity outcomes relating to continuous integration
  in github.
\newblock In {\em Proceedings of the 2015 10th Joint Meeting on Foundations of
  Software Engineering}, ESEC/FSE 2015, pages 805--816. ACM, 2015.

\bibitem{Wang2014GSE}
H.~Wang, Y.~Wang, and J.~Wang.
\newblock A participant recruitment framework for crowdsourcing based software
  requirement acquisition.
\newblock In {\em 2014 IEEE 9th International Conference on Global Software
  Engineering}, pages 65--73, 2014.

\end{thebibliography}

\end{document}